\documentclass[12pt]{article}
\usepackage{graphicx}
\usepackage{amssymb}
\usepackage{amsmath}

\newcommand{\bear}{\begin{array}}  
\newcommand {\eear}{\end{array}}
\newcommand{\bea}{\begin{eqnarray}}   
\newcommand{\eea}{\end{eqnarray}}
\newcommand{\beq}{\begin{eqnarray}}   
\newcommand{\eeq}{\end{eqnarray}}
\newcommand{\bef}{\begin{figure}}  \newcommand 
{\eef}{\end{figure}}
\newcommand{\bec}{\begin{center}}  \newcommand 
{\eec}{\end{center}}

\newcommand{\oneui}{\overline{\tilde{\chi}^0}}

\newcommand{\Slash}[1]{{\ooalign{\hfil/\hfil\crcr$#1$}}}

\begin{document}

\begin{titlepage}

\begin{flushright}
IPMU~11-0046 \\
ICRR-Report-583-2010-16\\
CALT 68-2824 \\
\end{flushright}

\vskip 1.35cm

\begin{center}

{\large 
{\bf Direct Detection of Electroweak-Interacting Dark Matter} 
}

\vskip 1.2cm

Junji Hisano$^{a,b}$,
Koji Ishiwata$^c$,
Natsumi Nagata$^{a,d}$,
and
Tomohiro Takesako$^e$ \\

\vskip 0.4cm

{ \it $^a$Department of Physics, 
Nagoya University, Nagoya 464-8602, Japan}\\
{\it $^b$Institute for the Physics and Mathematics of the Universe,
University of Tokyo, Kashiwa 277-8568, Japan}\\
{\it $^c$California Institute of Technology, Pasadena, CA 91125, USA}\\
{\it $^d$Department of Physics, 
University of Tokyo, Tokyo 113-0033, Japan}\\
{ \it $^e$Institute for Cosmic Ray Research,
University of Tokyo, Kashiwa 277-8582, Japan}
\date{\today}

\begin{abstract} 
  Assuming that the lightest neutral component in an $SU(2)_L$ gauge
  multiplet is the main ingredient of dark matter in the universe, we
  calculate the elastic scattering cross section of the dark matter
  with nucleon, which is an important quantity for the direct
  detection experiments.  When the dark matter is a real scalar or a
  Majorana fermion which has only electroweak gauge interactions, the
  scattering with quarks and gluon are induced through one- and
  two-loop quantum processes, respectively, and both of them give rise
  to comparable contributions to the elastic scattering cross section.
  We evaluate all of the contributions at the leading order and find
  that there is an accidental cancellation among them. As a result,
  the spin-independent cross section is found to be
  $\mathcal{O}(10^{-(46-48)}) ~{\rm cm}^2$, which is far below the
  current experimental bounds.
\end{abstract}

\end{center}
\end{titlepage}

\section{Introduction}
\label{sec:intro}

The existence of dark matter (DM) is one of the mysteries of the
universe. Its energy density in the universe, which is about six times
larger than that of baryon \cite{Komatsu:2010fb}, cannot be explained
in the standard model (SM) in particle physics.  Weakly interacting
massive particles (WIMPs) beyond the standard model are regarded as
promising candidates for the DM. If they have TeV-scale masses, their
relic abundance in the thermal history of universe may naturally
account for the observed value. For the past years, a lot of efforts
have been dedicated to the direct detection of WIMP DM, and their
sensitivities have been improving.  The XENON100 experiment
\cite{Aprile:2010um}, for example, has already started and announced
its first result, which gives a stringent constraint on the
spin-independent (SI) WIMP-nucleon elastic scattering cross section
$\sigma_N^{\text{SI}}$ ( $\sigma_N^{\text{SI}} < 2.4 \times
10^{-44}~{\rm cm}^2$ for WIMPs with a mass 50~GeV
\cite{Aprile:2011hx}).  Furthermore, ton-scale detectors for the
direct detection experiments are now planned and expected to have
significantly improved sensitivities.

Introduction of new particles with masses TeV scale is one of the
simplest extensions of the SM in order to explain the DM in the
universe if they are an electrically neutral component in an $SU(2)_L$
gauge multiplet ({\it $n$-tuplet}) with the hypercharge $Y$.  We call
them electroweak-interacting massive particles (EW-IMPs) in this
article.  EW-IMPs are assumed to interact with quarks and leptons only
via weak gauge interactions.  Neutral Wino and Higgsino in the minimal
supersymmetric standard model (MSSM) are examples of such particles
when the mixing with other states is negligible and squarks and
sleptons are heavy enough.  There is an earlier work
\cite{Drees:1996pk} which studied the direct detection of
Higgsino-like DM.  They gave the cross section of Higgsino-like DM
with nucleon by pointing out large loop contribution to Yukawa
couplings.  The cross sections of Wino/Higgsino DM with nucleon were
calculated in Ref.~\cite{Hisano:2004pv}.  In their study, they took
into account electroweak loop corrections and showed that the SI
interaction does not vanish in the large DM mass limit, which is a
distinctive feature of EW-IMP DM. There are several other articles
which give theoretical prediction for the cross section of EW-IMP
DM. The authors in Ref.~\cite{MDM} intensively studied EW-IMP DM (they
refer to it as Minimal dark matter), and a similar calculation was
also performed in Ref.~\cite{Essig:2007az}. There is, however,
inconsistency among Refs.~\cite{Hisano:2004pv,MDM,Essig:2007az}. In
addition, interaction of gluon with EW-IMP, which is also one of the
leading contributions in the EW-IMP DM-nucleon scattering, was
neglected in the references. This point was first pointed out in
Ref.~\cite{Hisano:2010fy}.

In this paper, we give accurate prediction for EW-IMP-nucleon
scattering in the direct detection experiments. Assuming that EW-IMPs
are the main component of the DM in the universe, we calculate the
cross section at the leading order of SM gauge couplings, including
the gluon contribution.  We provide the complete formulae for the
EW-IMP DM-nucleon scattering cross section in a general form, and show
that the SI cross section is below the current experimental bounds in
all the cases we have studied. We also give numerical results for
spin-dependent (SD) cross section.

This paper is organized as follows: in Section~\ref{sec:model} we
explain the EW-IMP DM scenario. In Section~\ref{sec:eff} general
formulae of the elastic scattering cross section of dark matter with
nucleon are summarized. In Section~\ref{sec:results} we derive the
cross section of the EW-IMP with nucleon and show the numerical
results.  Section~\ref{sec:conclusion} is devoted to conclusion.

\section{EW-IMP Dark Matter}
\label{sec:model}

As we described in the Introduction, EW-IMPs are an electrically
neutral component in $n$-tuplet of $SU(2)_L$ with the hypercharge $Y$.
In this section, we explain characteristics of EW-IMP DM assuming the
EW-IMPs are fermionic.  Fermionic EW-IMPs are popular, {\it e.g.}, Wino
or Higgsino in the MSSM.  We will discuss scalar EW-IMPs in the end of
this section.

When the fermionic $n$-tuplet has only $SU(2)_L\times U(1)_Y$ gauge
interactions in the SM and all the components have a common mass at
tree level, the charged components become heavier than the neutral one
because of quantum loop corrections \cite{MDM}. The typical value of
the mass difference $\Delta M$ is $ {\cal O}(100)~{\rm MeV}$. Thus,
the neutral component of the multiplet could be a DM candidate in the
universe if it is stable.  It is found that fermionic EW-IMPs with $n
\geq 5$ could be stable without imposing a certain symmetry to forbid
its decay, such as $R$ parity in the MSSM, since the gauge and Lorentz
invariance prevent the DM candidate from decaying via the
renormalizable interactions, and they turn out to be stable even if
non-renormalizable effective operators are considered under the
cut-off scale as large as the GUT-scale or Planck-scale~\cite{MDM,
  MDM2}\footnote{
In this article we implicitly suppose a symmetry to
prevent EW-IMPs from decaying for $2\le n < 5$.
}.

The thermal relic abundance could also explain the observed DM energy
density when the EW-IMP mass is over $1~{\rm TeV}$ for $n\ge 2$.  In
Ref.~\cite{Hisano:2006nn}, the thermal relic abundance of Wino in the
MSSM (which is EW-IMP with $n=3$ and $Y=0$) is evaluated.\footnote
{In the evaluation of thermal relic abundance of EW-IMPs
with $n>2$, the Sommerfeld effect in the annihilation cross section
of EW-IMP \cite{Hisano:2003ec,Hisano:2004ds} should be included
\cite{Hisano:2006nn}.
}
The thermal relic abundances for EW-IMPs with $n=2$ and $5$ ($Y=0$) are
also evaluated \cite{MDM2}. According to those studies, the EW-IMP mass
$1,~2.7,$ and $10~{\rm TeV}$ is suggested in order to explain the
observed DM abundance when $n=2,~3,$ and $5$, respectively, and $Y=0$.

Now we explicitly show the Lagrangian of EW-IMP DM scenario for our
calculation.  In the case of the multiplet with the hypercharge $Y =
0$, the following Lagrangian is introduced to the SM :
\begin{equation}
\begin{split}
\Delta \mathcal{L} 
&= \frac{1}{2} \overline{\tilde \chi} ( i \Slash{D} - M_0 )
 \tilde \chi~, \\ 
\end{split}
\label{yzerolagrangian}
\end{equation}
where $\tilde \chi$ is an $SU(2)_L$ $n$-tuplet fermion, and
$D_{\mu}=\partial_{\mu}-ig_1 Y B_{\mu}-ig_2 W^{a}_{\mu}T^{a}$
($a=1,2,3$) is the gauge covariant derivative. Here $g_1$ and $g_2$ is
$U(1)_Y$ and $SU(2)_L$ gauge coupling constants, respectively, and
$T^a$ is for the generators of $SU(2)_L$ gauge group. The tree-level
mass of the multiplet is denoted as $M_0$.  The neutral component of
the $n$-tuplet is a Majorana fermion, while the other charged
components are Dirac fermions. The neutral component has no
interaction with gauge bosons by itself so that the elastic scattering
of EW-IMP with nucleon is induced via loop diagrams.

On the other hand, for $Y \neq 0$ $n$-tuplet, the following term is
added to the SM Lagrangian,
\begin{equation}
\begin{split}
\Delta \mathcal{L} 
&= \overline{\tilde \psi} ( i \Slash{D} - M_0 ) \tilde \psi~.
\end{split}
\label{deltal}
\end{equation}
Here, $\tilde \psi$ is an $SU(2)_L$ $n$-tuplet Dirac fermion so that
the fermion have an $SU(2)_L\times U(1)_Y$ invariant mass term. Note
that Dirac fermions as the DM in the universe have been severely
constrained by the direct detection experiments because of its large
elastic scattering cross section via coherent vector interaction with
target nuclei.

The situation changes if introducing the extra effective operators
which give rise to a mass splitting $\delta m$ between the left- and
right-handed components of the neutral Dirac fermion after the
electroweak symmetry breaking. In this case, the neutral component is
no longer mass eigenstate; the mass splitting decomposes the
neutral Dirac fermion, $\tilde \psi^0$, into two Majorana fermions,
$\tilde \chi^0$ and $\tilde \eta^0$ :
\begin{equation}
\begin{split}
\tilde \psi^0 = \frac{1}{\sqrt{2}} (\tilde \chi^0 + i \tilde \eta^0 ) .
\end{split}
\label{deltal2}
\end{equation}
Hereafter we treat $\tilde \chi^0$ as the lighter one, without loss of
generality. Now the lighter component could be the DM candidate. It
does not have any vector interaction by itself so that the DM-nucleon
elastic scattering by $Z$-boson exchange is forbidden. In addition, if
the mass splitting is large enough ($\delta m \gg \mathcal{O} (10)~
\mathrm{keV}$), the DM-nucleon {\it inelastic} scattering is
suppressed kinematically.  Thus, the EW-IMPs avoid the constraints from
the direct detection experiments. On the other hand, large $\delta m$
may also induce sizable contribution of the Higgs-boson exchange to
the DM-nucleon scattering at tree level. To avoid all these
constraints, in this article, we simply assume $\mathcal{O} (10
)~\mathrm{keV} \ll \delta m \ll \Delta M$ so that the elastic
scattering of EW-IMP with nucleon is dominated by loop diagrams.

In the following discussion, we consider the scenarios with Majorana
fermion EW-IMPs, with either $Y=0$ or $Y\neq 0$. We simply express it
as $\tilde \chi^0$ and the charged Dirac components which couple with
$\tilde{\chi}^0$ as $\tilde{\psi}^\pm$.

It is straightforward to extend our calculation to scalar EW-IMPs.
Contrary to the fermionic EW-IMPs, the scalar EW-IMPs have
renormalizable self-interactions with Higgs boson, which contribute to
the SI cross section of scalar EW-IMP with nucleon.  When the
interaction is suppressed enough to be ignored, elastic scattering of
the EW-IMP with nucleon is induced by loop diagrams. In that case, the
SI cross section of scalar EW-IMP with nucleon should agree with that
of the fermionic EW-IMP when the EW-IMP mass is much larger than weak
gauge boson masses.  This is because the Lagrangians for both of
fermionic and scalar EW-IMPs are the same in the non-relativistic limit
or at the leading of the velocity expansion.  Needless to say, the SD
cross section of scalar EW-IMP with nucleon vanishes since the scalar
EW-IMP has spin zero.

\section{Effective Interactions for DM-Nucleon Scattering}
\label{sec:eff}

In this section, we briefly review evaluation of the elastic
scattering cross section of DM with nucleon, which is an important
quantity for the direct detection experiments, assuming the DM is a
Majorana fermion.  The formulation given here is originally derived in
Ref.~\cite{Drees:1993bu}.  See also Refs.~\cite{Hisano:2010ct,
  Jungman:1995df} for further details.

First, we write down the effective interactions of DM with
light quarks ($q=u,d,s$) and gluon\footnote
{ We only keep the operators which give rise to the leading
  contributions for the DM-nucleon scattering with the
  non-relativistic velocity.  Moreover, in order to remove the
  redundant terms, we use the integration by parts and the classical
  equation of motion for the operators when we construct the
  low-energy effective Lagrangian~\cite{Politzer}. }:
\begin{eqnarray}
{\cal L}^{\rm eff}&=&\sum_{q=u,d,s}{\cal L}^{\rm{eff}}_q +{\cal
 L}^{\rm{eff}}_g \ , 
\label{eq:eff} \\
{\cal L}^{\rm{eff}}_q
&=& 
d_q\ \oneui\gamma^{\mu}\gamma_5\tilde{\chi}^0\  \bar{q}\gamma_{\mu}\gamma_5 q
+
f_q m_q\ \oneui\tilde{\chi}^0\ \bar{q}q 
\cr
&+& \frac{g^{(1)}_q}{M} \ \oneui i \partial^{\mu}\gamma^{\nu} 
\tilde{\chi}^0 \ {\cal O}_{\mu\nu}^q
+ \frac{g^{(2)}_q}{M^2}\
\oneui(i \partial^{\mu})(i \partial^{\nu})
\tilde{\chi}^0 \ {\cal O}_{\mu\nu}^q \
,
\label{eff_lagq}
\\
{\cal L}^{\rm eff}_{ g}&=&
f_G\ \oneui\tilde{\chi}^0 G_{\mu\nu}^aG^{a\mu\nu}
 \ .
\label{eff_lagg}
\end{eqnarray}
Here, $M$ and $m_q$ are the masses of the DM and quarks, respectively.
The field strength tensor of the gluon field is denoted by $G^a_{\mu
  \nu}$. The last two terms in Eq.~(\ref{eff_lagq}) include the quark
twist-2 operator, ${\cal O}_{\mu\nu}^q$, which is defined as,
\beq {\cal O}_{\mu\nu}^q&\equiv&\frac12 \bar{q} i
\left(D_{\mu}\gamma_{\nu} +
  D_{\nu}\gamma_{\mu} -\frac{1}{2}g_{\mu\nu}\Slash{D}
\right) q \ .
\eeq
The coefficients of the operators in Eqs.~(\ref{eff_lagq}) and
(\ref{eff_lagg}) are to be determined in the following section.

The cross section of DM with nucleon ($N=p,n$) is calculated from the
effective Lagrangian in terms of SI and SD effective couplings, $f_N$
and $a_N$,
\begin{eqnarray}
  \sigma_N&=&
  \frac{4}{\pi}m_R^2
  \left[\left| f_N\right|^2+3\left|a_N\right|^2\right]\ , 
\label{sigma}
\end{eqnarray}
where $m_R\equiv M m_N/(M+m_N)$ ($m_N$ is nucleon mass). The SI and SD
effective couplings are given as scattering matrix element of the
effective Lagrangian between initial and final states. The results are
\begin{eqnarray}
f_N/m_N&=&\sum_{q=u,d,s}
f_q f_{Tq}
+\sum_{q=u,d,s,c,b}
\frac{3}{4} \left(q(2)+\bar{q}(2)\right)\left(g_q^{(1)}+g_q^{(2)}\right)
-\frac{8\pi}{9\alpha_s}f_{TG} f_G 
~,
\label{f}
\\
a_{N}&=&\sum_{q=u,d,s} d_q \Delta q_N \ ,
\label{an}
\end{eqnarray}
where 
\begin{eqnarray}
&& \langle N \vert m_q \bar{q} q \vert N\rangle/m_N = f_{Tq}\ , \ 
f_{TG}= 1-\sum_{u,d,s}f_{Tq} \ ,
\label{ftq}
\\
&& \langle N(p)\vert 
{\cal O}_{\mu\nu}^q
\vert N(p) \rangle 
=\frac{1}{m_N}
(p_{\mu}p_{\nu}-\frac{1}{4}m^2_N g_{\mu\nu})\
(q(2)+\bar{q}(2)) \ ,
\label{nucleon_matrices} \\
&& \langle N \vert 
\bar{q}\gamma_{\mu}\gamma_5  q \vert N \rangle = 2 s_{\mu}\Delta q_N \ .
\label{sfra}
\end{eqnarray}
Here $\alpha_s\equiv g_s^2/4\pi$ ($g_s$ is the coupling constant of
$SU(3)_C$), $q(2)$ and $\bar{q}(2)$ are the second moments of the
parton distribution functions (PDFs), and $s_{\mu}$ is the spin of the
nucleon.  Note the factor $1/ \alpha_s$ in front of $f_G$ in
Eq.~(\ref{f}). It makes the gluon contribution comparable to the
light-quark contribution, despite the interactions of DM with gluon
are induced by higher-loop processes than those with light quarks
\cite{Hisano:2010fy}. In the present case, as discussed in the
previous section, the DM-quark tree-level scattering is highly
suppressed, and thus the one-loop processes become dominant for the
DM-light quark effective interactions.  Therefore, in order to give
accurate prediction of the scattering cross section, we should
evaluate not only the one-loop diagrams with light quarks but also the
two-loop diagrams with gluon.

In Table~1 we list the numerical values for the parameters that we
used in this article.  The mass fractions of light quarks, $f_{Tq}$
$(q=u,d,s)$, are calculated using the results in
Refs.~\cite{Corsetti:2000yq, Ohki:2008ff, Cheng:1988im}.  The
procedure for evaluating them is described in these references and
Ref.~\cite{Hisano:2010ct}.  The second moments of PDFs of quarks and
anti-quarks are calculated at the scale $\mu=m_Z$ ($m_Z$ is $Z$-boson
mass) using the CTEQ parton distribution
\cite{Pumplin:2002vw}\footnote
{As will be described later, the terms with quark twist-2 operators in
  Eq.~(\ref{eff_lagq}) are induced by the one-loop diagrams in which
  the loop momentum around the weak-boson mass scale yields dominant
  contribution. This fact verifies the use of the second moments of
  PDFs at the $m_Z$ scale. See also discussion relevant to it in
  Ref.~\cite{Hisano:2010yh}. }.
Finally, the spin fractions, $\Delta q_N$, in Eq.~(\ref{sfra}) are
obtained from Ref.~\cite{Adams:1995ufa}. The second moments and the
spin fractions for neutron are to be obtained by exchanging the values
of up quark for those of down quark.
\begin{table}
\label{table1}
\begin{center}
\begin{tabular}{|l|l|}
\hline
\multicolumn{2}{|c|}{Mass fraction}\cr
\hline
\multicolumn{2}{|c|}{(for proton) }\cr
\hline 
$f_{Tu}$& 0.023\cr
$f_{Td}$& 0.032\cr
$f_{Ts}$&0.020\cr
\hline
\multicolumn{2}{|c|}{(for neutron)}\cr
\hline
$f_{Tu}$&0.017\cr
$f_{Td}$& 0.041\cr
$f_{Ts}$& 0.020 \cr
\hline
\end{tabular}
\hskip 1cm 
\begin{tabular}{|l|l||l|l|}
\hline
\multicolumn{4}{|c|}{Second moment at $\mu=m_Z$ }\cr
\multicolumn{4}{|r|}{(for proton) }\cr
\hline
$u(2)$&0.22&$\bar{u}(2)$& 0.034\cr
$d(2)$&0.11&$\bar{d}(2)$&0.036\cr
$s(2)$&0.026&$\bar{s}(2)$&0.026\cr
$c(2)$&0.019&$\bar{c}(2)$&0.019\cr
$b(2)$&0.012&$\bar{b}(2)$&0.012\cr
\hline
\end{tabular}
\hskip 1cm 
\begin{tabular}{|l|l|}
\hline
\multicolumn{2}{|c|}{Spin fraction }\cr
\multicolumn{2}{|r|}{(for proton)}\cr
\hline
$\Delta u_p$& 0.77\cr
$\Delta d_p$& -0.49\cr
$\Delta s_p$& -0.15\cr
\hline 
\end{tabular}
\end{center}
\caption{Parameters for quark and gluon matrix elements used in this
  paper.} 
\end{table}
%

\section{Results}
\label{sec:results}

\subsection{Coefficients of the Effective Lagrangian}
Now we evaluate the coefficients of the effective interactions
displayed in Eqs.~(\ref{eff_lagq}) and (\ref{eff_lagg}).  From
Eqs.~(\ref{deltal}) and (\ref{deltal2}), it is found that the EW-IMP,
${\tilde \chi^0}$, interacts with the weak gauge bosons
($W^{\pm}_{\mu}, Z^0_{\mu}$) as
\begin{equation}
\begin{split}
\Delta \mathcal{L}_{\rm{int.}}
&= \left[\frac{g_2}{4} \sqrt{n^2 -(2 Y +1)^2}~\overline{\tilde \chi^0}
 \gamma^{\mu} \tilde \psi^-~W_{\mu}^+   
+ \frac{g_2}{4}  \sqrt{n^2 -(2 Y - 1)^2}~
 \overline{\tilde \chi^0} \gamma^{\mu} \tilde \psi^+ ~ W_{\mu}^- +
 {\rm h.c.} \right]\\ 
&~~ + \frac{i g_2(- Y )}{\mathrm{cos}\theta_W} ~ \overline{\tilde
 \chi^0} \gamma^{\mu} \tilde \eta^0~Z_{\mu}^0, 
\end{split}
\label{lag_eff}
\end{equation}
where $\theta_W$ is the weak mixing angle. The Majorana field $\tilde
\eta^0$ is introduced for the cases of $Y\ne 0$. (See
Eq.~(\ref{deltal2}).)  In either case ($Y=0$ or $Y \neq 0$), the EW-IMP
does not have any interaction by itself.  Thus, it is loop diagrams
that yield the leading contribution to the EW-IMP-nucleon elastic
scattering cross section.

\begin{figure}[t]
\begin{center}
\includegraphics[width=100mm]{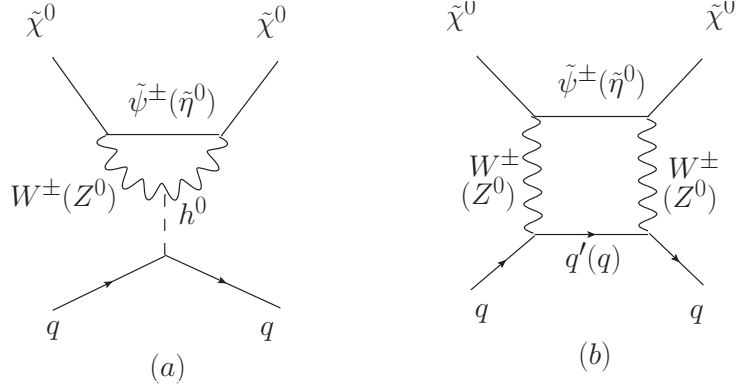}
\caption{One-loop diagrams which induce effective interactions of
  EW-IMP DM with light quarks.  There are also $W$-($Z$-) boson
  crossing diagrams, which are not shown here.  }
\label{fig:1-loop-diagramsWZ}
\end{center}
\end{figure}

\begin{figure}[t]
\begin{center}
\includegraphics[width=100mm]{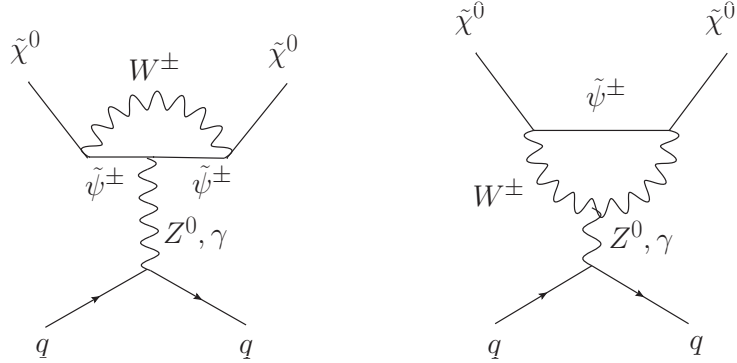}
\caption{One-loop diagrams which correspond to the one-loop quantum
  correction to the EW-IMP-$Z$ ($\gamma$) interaction vertex. These
  contributions turn out to vanish.}
\label{fig:vanish}
\end{center}
\end{figure}

First, we consider the one-loop processes.  The relevant diagrams are
shown in Figs.~\ref{fig:1-loop-diagramsWZ} and \ref{fig:vanish}.
%
%
%
The diagrams in Fig.~\ref{fig:1-loop-diagramsWZ} give
rise to the coefficients in Eq.~(\ref{eff_lagq}) as
\begin{equation}
\begin{split}
f_q
&= \frac{\alpha_2^2}{4 m_{h}^2}\left[
\frac{n^2-(4Y^2+1)}{8m_W}g_{\rm H}(w)+
\frac{Y^2}{4m_Z\mathrm{cos}^4 \theta_W }  g_{\rm H} (z) \right]+  \frac{\left(
 (a^V_q)^2 - (a^A_q)^2 \right) 
 Y^2}{\mathrm{cos}^4 \theta_W} \frac{\alpha_2^2}{m_Z^3}
 g_{\rm S}(z), 
\label{eq:higgs}
\end{split}
\end{equation}
\begin{equation}\label{eq:axial-Z}
\begin{split}
d_q
&= \frac{n^2-(4Y^2+1)}{8}\frac{\alpha^2_2}{m^2_W}g_{\rm AV}(w)+
\frac{2 \left( (a^V_q)^2 + (a^A_q)^2 \right) Y^2}{\mathrm{cos}^4
 \theta_W} \frac{\alpha_2^2}{m_Z^2} g_{\text{AV}}(z), 
\end{split}
\end{equation}
\begin{equation}
\begin{split}
&g_q^{(1) } = \frac{n^2-(4Y^2+1)}{8}\frac{\alpha^2_2}{m^3_W} g_{\rm T1}(w)+
\frac{2 \left( (a^V_q)^2 + (a^A_q)^2 \right)
 Y^2}{\mathrm{cos}^4 \theta_W} \frac{\alpha_2^2}{m_Z^3} g_{\rm T 1}(z) , 
\label{eq:twist2-1}
\end{split}
\end{equation}
\begin{equation}
\begin{split} 
&g_q^{(2) } = \frac{n^2-(4Y^2+1)}{8}\frac{\alpha^2_2}{m^3_W} g_{\rm T2}(w)+
\frac{2 \left( (a^V_q)^2 + (a^A_q)^2 \right)
 Y^2}{\mathrm{cos}^4 \theta_W} \frac{\alpha_2^2}{m_Z^3} g_{\rm T 2}(z).
\label{eq:twist2-2}
\end{split}
\end{equation}
Here, $m_{h}$ and $m_W$ are the masses of Higgs and $W$ bosons,
respectively, and $\alpha_2 = g_2^2/4 \pi$.  We also define the vector
and axial-vector couplings of quarks with $Z$ boson as
\begin{equation}
a^V_q = \frac{1}{2} T_{3 q} - Q_q ~\mathrm{sin}^2 \theta_W , ~~~~~~
a^A_q = - \frac{1}{2} T_{3 q} ,
\end{equation}
where $T_{3 q}$ and $Q_q$ denote the weak isospin and the charge of
quark $q$, respectively. Furthermore, we parametrize $w\equiv
m^2_W/M^2$ and $z\equiv m^2_Z/M^2$ in the above expressions.  The
first term in Eq.~(\ref{eq:higgs}) is induced by the Higgs-boson
exchange process, shown in diagram (a) of
Fig.~\ref{fig:1-loop-diagramsWZ}.  The other terms in
Eqs.~(\ref{eq:higgs}-\ref{eq:twist2-2}) are all obtained from diagram
(b).  The mass functions, $g_H(x), g_S(x),g_{\rm{AV}}(x), g_{T 1}(x)$,
and $g_{T2}(x)$, in Eqs.~(\ref{eq:higgs}-\ref{eq:twist2-2}) are given
in Appendix~\ref{app1}. We ignored the mass differences between
${\tilde \chi^0}$ and ${\tilde \psi^{\pm}}$ and also between ${\tilde
  \chi^0}$ and ${\tilde \eta^0}$ here. The loop integrals for the
diagrams are finite, and they are dominated by the loop momentum
around the weak-boson mass scale. We note here that some of these mass
functions do not vanish in the limit of $w,z\rightarrow 0$
\cite{Hisano:2004pv}. This implies that the SI interactions of EW-IMP
with light quarks are not suppressed even if the EW-IMP mass is much
larger than the gauge boson masses as we described in the
Introduction.  On the other hand, we found that both diagrams in
Fig.~\ref{fig:vanish} vanish separately by explicit calculation.

\begin{figure}[t]
\begin{center}
\includegraphics[width=150mm]{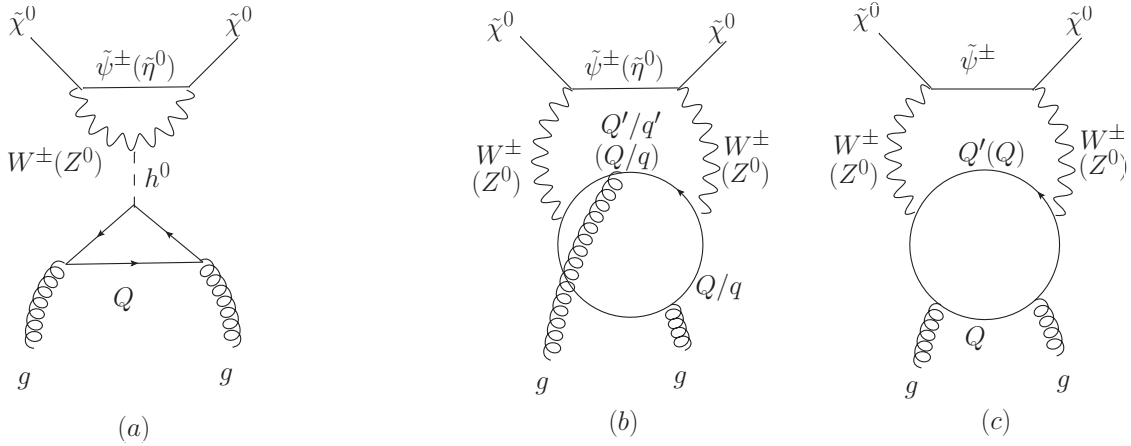}
\caption{Relevant two-loop diagrams which contribute to effective
  scalar coupling of EW-IMP DM with gluon.There are also $W$-($Z$-)
  boson crossing diagrams, which are not shown here.}
\label{fig:2-loop-diagramsWZ}
\end{center}
\end{figure}

Next, let us move to the two-loop diagrams which yield the effective
scalar coupling of EW-IMP with gluon, $f_G$.  These diagrams are
presented in Fig.~\ref{fig:2-loop-diagramsWZ}.  For each diagram, the
gluon contributions are classified into two types in terms of the
momentum scale which dominates in the loop integration. So-called
``short-distance'' contribution means that the momentum is typically
masses of heavy particles, such as DM particle or the weak gauge
bosons in the diagrams, and ``long-distance'' one means that the
momentum is typically mass of quark in the loop diagrams. Among the
latter one, the contributions in which the light quarks run are
already incorporated in the mass fractions $f_{T_q}$ defined in
Eq.~(\ref{ftq}). Therefore, we do not need to add them in the
calculation of the gluon contribution; otherwise we would doubly count
them \cite{Hisano:2010ct}.  Consequently, the gluon contribution from
each diagram is given by
\begin{equation}
\begin{split}
f^{(i)}_G 
&= \sum_{q=u,d,s,c,b,t} f^{(i)}_G |_q^{\text{SD}} + \sum_{Q = c,b,t} c_Q
 f^{(i)}_G |_Q^{\text{LD}} .
\end{split}
\end{equation}
Here $f^{(i)}_G |_q^{\text{SD}}$ and $f^{(i)}_G |_q^{\text{LD}}$
denote the short-distance and long-distance contributions of quark $q$
in the loop in diagram $(i)$ ($i=a,b,c$) of
Fig.~\ref{fig:2-loop-diagramsWZ}, respectively. We also take into
account large QCD corrections in the long-distance contributions
\cite{Djouadi2000} by using $c_Q = 1 + 11 \alpha_s (m_Q) / 4 \pi$ ($Q
=c, b, t$). We take $c_c = 1.32, c_b = 1.19$, and $c_t = 1$ for
$\alpha_s (m_Z) = 0.118$ in our calculation.  Note that the
long-distance contribution is gauge invariant. This is because its
contribution to the operator $f_GG^a_{\mu\nu}G^{a\mu\nu}$ is evaluated
from scalar-type effective operator
$f_qm_q\bar{q}q$~\cite{Drees:1993bu,Hisano:2010ct}. (See also later
discussion where the explicit calculations are given.) Consequently,
the gauge invariance of the short-distance contribution is guaranteed
since summation of the both contribution is obviously gauge invariant.
Then, the effective coupling of EW-IMP with gluon is obtained as
\begin{equation}
 f_G=f_G^{(a)}+f_G^{(b)}+f_G^{(c)}~.
\end{equation}

Let us see each diagram closely.  It is obvious that diagram (a) gives
the long-distance contribution.  Thus, we sum up for heavy quarks in
the loop, and get
\begin{equation}
 f^{(a)}_G=-\frac{\alpha_s}{12\pi}\times 
\frac{\alpha_2^2}{4 m_{h}^2}
\sum_{Q=c,b,t}c_Q
\left[
\frac{n^2-(4Y^2+1)}{8m_W}g_{\rm H}(w)+
\frac{Y^2}{4m_Z\mathrm{cos}^4 \theta_W }  g_{\rm H} (z) \right]~.
\label{fga}
\end{equation}
Here the first and second terms in the bracket come from $W$- and
$Z$-boson exchanges, respectively.  As we described above, this
long-distance contribution is given by effective scalar-type coupling
(of the Higgs contribution) as $-\frac{{\alpha}_s}{12\pi} f_Q$.

For the calculation of diagrams (b) and (c), on the other hand, we
follow the steps supplied in Ref.~\cite{Hisano:2010ct}. In the work,
the systematic calculation for the $W$-boson exchange diagrams at
two-loop level in the Wino DM scenario is given. The procedure is
applicable to compute the two-loop diagrams in our case.  For the
$W$-boson exchange diagrams, the analytic result is simply given by
multiplying a factor $({n^2-(4Y^2+1)})/{8}$ to their result of the
Wino DM case.  (as the first term in Eq.~(\ref{fga})).  Even for the
$Z$-boson exchange contribution, the calculation is straightforward
extension.  In the following we outline the calculation and give the
result of the $Z$-boson exchange contribution.

First, we calculate the vacuum polarization tensor of $Z$ boson in the
gluon background field, $\Pi^Z_{\mu \nu} (p)$. It is given as
\begin{equation}
\begin{split}
&i \Pi^Z_{\mu \nu} (p) \\
&= -\sum_{[q]} \int \frac{\mathrm{d}^4 l}{(2 \pi)^4} \frac{g_2^2}{\cos^2
 \theta_W}~\mathrm{Tr}_{\rm{L+C}} \{  \gamma_{\mu} (a^V_q + a^A_q
 \gamma_5 ) S_q(l)  \gamma_{\nu}  (a^V_q + a^A_q \gamma_5 ) \tilde
 S_q(l- p)  \} , 
\end{split}
\end{equation}
where $\rm{Tr}_{\rm{L+C}}$ denotes the trace over the Lorentz and
color indices, and $[q]$ means the sum is taken over all quarks for
the short-distance contribution and heavy quarks for the long-distance
contribution.  The quark propagators $S_q(p)$ and $\tilde S_q(p)$ are
under the gluon background field with the Fock-Schwinger gauge. (See
Appendix A in Ref.~\cite{Hisano:2010ct}.)  We decompose the
polarization function as
\begin{equation}
\begin{split}
\Pi^Z_{\mu \nu} (p) =  \left( - g_{\mu \nu} + \frac{p_{\mu} p_{\nu}}{
 p^2} \right)~ \Pi^Z_T (p^2) + \frac{p_{\mu} p_{\nu}}{p^2}~ \Pi^Z_L
 (p^2). 
\end{split}
\label{polz}
\end{equation}
By the explicit calculation, we found that the longitudinal part
$\Pi_L (p^2)$ does not contribute to $f_G$\footnote{
  This is consistent with the fact that EW-IMPs have no tree-level
  coupling with Higgs boson and Goldstone bosons of the electroweak
  symmetry breaking, which turn into the longitudinal modes of $Z$ and
  $W$ bosons.
}.  Thus, we only calculate transverse component, $\Pi^Z_T (p^2)$.
The contributions from diagrams (b) and (c) to $\Pi^Z_T (p^2)$ are
given as
\begin{eqnarray}
 \left[ \Pi^Z_T (p^2) \right]_{(b)}^q|_{GG} &=& \frac{\alpha_2
  \alpha_s}{6 \cos^2 \theta_W}(G^a_{\mu \nu})^2 \left[ (a^V_q)^2
+(a^A_q)^2   \right] \left[p^2 (B^{(2,2)}_{1}+B^{(2,2)}_{21})+6
B^{(2,2)}_{22} \right], \nonumber \\
 \left[ \Pi^Z_T (p^2) \right]_{(c)}^q|_{GG} &=& \frac{2\alpha_2
  \alpha_s}{\cos^2 \theta_W}(G^a_{\mu \nu})^2 m^2_q\bigl[ \{(a^V_q)^2
+(a^A_q)^2\}(-p^2B^{(4,1)}_1+2B^{(4,1)}_{22})
\nonumber \\
&& ~~~~~~~~~~~~~~~~~~~~~~~~~~~~~~~~~~~~~~~~~
 -2(a^A_q)^2(p^2
B^{(4,1)}_{21}+4B^{(4,1)}_{22})\bigr]  ,
\label{pol_bc}
\end{eqnarray}
where $\left[ \Pi^Z_T (p^2) \right]_{(i)}^q$ denotes the contribution
from diagram $(i)$ ($i = b, c$) with quark $q$ running in the loop.
Here, ``$|_{GG}$'' represents that we keep only the terms which are
proportional to $G^a_{\mu\nu}G^{a\mu\nu}$ in the polarization
function, since the terms proportional to $G^a_{\mu\nu}G^{a\mu\nu}$
contribute to $f_G$. The loop functions in Eq.~(\ref{pol_bc}) are
defined as follows:
\begin{eqnarray}
&& p^{\mu}B_1^{(n,m)}\equiv\int\frac{d^4 k}{i\pi^2}\frac{k^{\mu}}
{[k^2-m_q^2]^n[(k+p)^2-m_q^2]^m}~,
\\ &&
p^{\mu}p^{\nu}B_{21}^{(n,m)}
+g^{\mu\nu}B_{22}^{(n,m)}
\equiv\int\frac{d^4
 k}{i\pi^2}\frac{k^{\mu}k^{\nu}} 
{[k^2-m_q^2]^n[(k+p)^2-m_q^2]^m}~.
\end{eqnarray}
As is the same in the calculation of two-loop $W$-boson exchange
diagrams discussed in Ref.~\cite{Hisano:2010ct}, the diagrams (b) and
(c) yield the short-distance contribution and long-distance
contribution, respectively.  We have checked this identification by
explicit calculation.  This is also confirmed from the fact that the
long distance-contribution corresponds to the one which is calculated
from quark triangle diagram using effective scalar-type coupling $f_q$
in the $M$ and $m_{W/Z} \rightarrow \infty$ limit
\cite{Hisano:2010ct}, as we mentioned before. Therefore, $\Pi^Z_T
(p^2)|_{GG}$ is obtained as
\begin{equation}
 \Pi^Z_T (p^2)|_{GG}=\sum_{q=u,d,s,c,b,t}\left[ \Pi^Z_T (p^2)
\right]_{(b)}^q|_{GG} +\sum_{Q=c,b,t}c_Q \left[ \Pi^Z_T (p^2)
\right]_{(c)}^Q|_{GG} .
\end{equation}
The results of $\left[ \Pi^Z_T (p^2) \right]_{(b)}^q|_{GG}$ and
$\left[ \Pi^Z_T (p^2) \right]_{(c)}^Q|_{GG}$ are given in Appendix~\ref{app2}.

With these $Z$-boson polarization functions, we get the contribution
from $Z$-boson exchange diagrams at two-loop level. Combining it with
the contribution from $W$-boson exchange diagrams, we eventually
obtain the effective coupling $f_G$ from diagrams (b) and (c) as
\begin{equation}
 \begin{split}
 f^{(b)}_G+f^{(c)}_G 
=\frac{\alpha_s\alpha^2_2}{4 \pi}\left[
\frac{n^2 - ( 4Y^2 +1)}{8m^3_W}g_{\rm W}(w,y)
+\frac{Y^2}{4m_Z^3 \mathrm{cos}^4 \theta_W}g_{\rm Z}(z,y)
\right]~,
 \end{split}
\label{fgbc}
\end{equation}
where $y \equiv m_t^2/ M^2$ ($m_t$ is the top quark mass), and the
mass functions, $g_{\rm W}(z,y)$ and $g_{\rm Z}(z,y)$, are given in
Appendix~\ref{app2}.  We ignore the mass of quarks except that of top
quark in this computation.

\subsection{Spin-independent and Spin-dependent  Cross Sections}
\begin{figure}[t]
\begin{center}
\includegraphics[width=100mm]{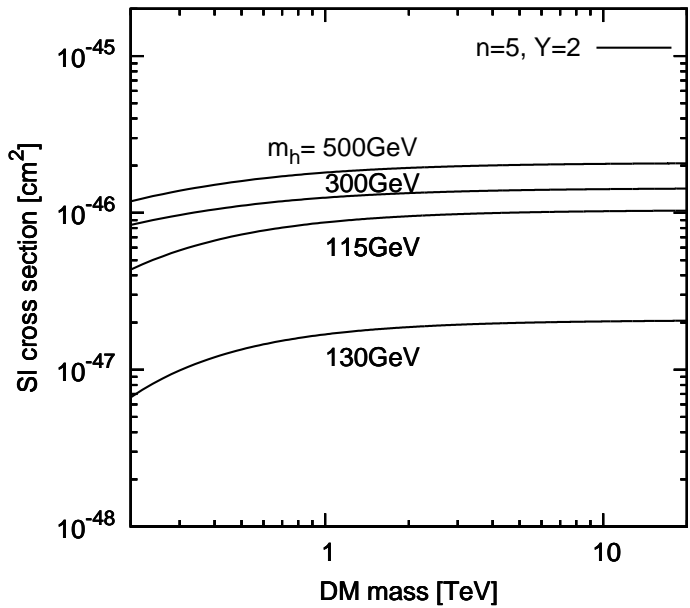}
\caption{DM-proton SI cross section for $n=5$ and $Y=2$. We take
  Higgs-boson mass as $m_{h}=$130, 115, 300 and 500~GeV from bottom to
  top in this figure.}
\label{fig:SIn5Y2}
\end{center}
\end{figure}

With the effective couplings derived above, we evaluate the cross
section of EW-IMP with nucleon. First, we discuss the SI cross section.
In order to look over the behavior of the cross section, we plot the
SI cross section as a function of the EW-IMP mass in
Fig.~\ref{fig:SIn5Y2}.  In this figure, we set $n=5$ and $Y=2$ as an
example of the case where both $W$- and $Z$-boson exchange diagrams
contribute, and take Higgs-boson mass as $m_{h}=$130, 115, 300 and
500~GeV from bottom to top. It is found that the SI cross section has
little dependence on the EW-IMP mass when the mass is larger than
${\mathcal O}(1)~{\rm TeV}$.  We have checked such a behavior for
other $n$ and $Y$ cases.  The cross section ranges from $10^{-47}~{\rm
  cm}^2$ to $10^{-46}~{\rm cm}^2$ in this figure when Higgs-boson mass
increased from $m_h=115~{\rm GeV}$ to $500~{\rm GeV}$. (Although it is
not monotonic.)

\begin{figure}
\begin{minipage}{0.5 \hsize}
\begin{center}
\includegraphics[width=80mm]{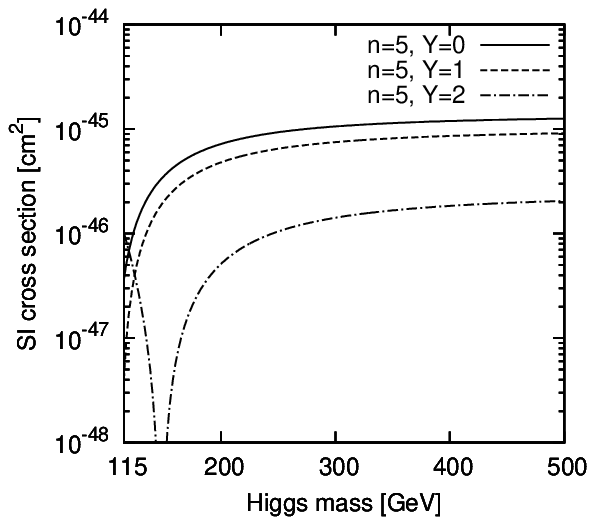}
\end{center}
\end{minipage}
\begin{minipage}{0.5 \hsize}
\begin{center}
\includegraphics[width=80mm]{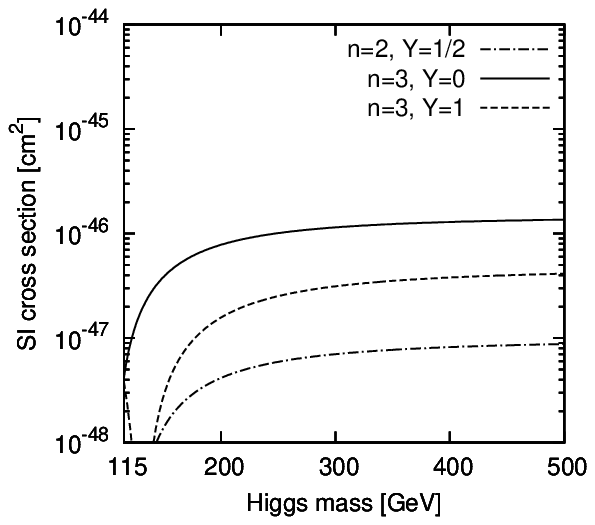}
\end{center}
\end{minipage}
\caption{SI cross sections of DM-proton elastic scattering as a
  function of Higgs-boson mass for $n=5$ (left panel) and $n=2, 3$ (right
  panel).  We take $M=1$, 2.7, 10 TeV for $n=$2, 3, and 5,
  respectively. In left panel, solid, dashed, and dash-dotted
  lines represent $n=5$ with $Y=0$, 1, and 2 cases, respectively.  In
  right panel $(n, Y)=(3, 0)$, $(3, 1)$, and $(2, 1/2)$ correspond
  to solid, dashed, and dash-dotted lines, respectively.}
\label{fig:SI}
\end{figure}

In Fig.~\ref{fig:SI}, we plot the EW-IMP-proton SI cross section as a
function of Higgs-boson mass for $n=5$ (left panel) and both $n=2$ and
$n=3$ (right panel).  The EW-IMP mass is taken to be equal to 1, 2.7,
and 10 TeV for $n=2$, 3, and 5, respectively\footnote{
  Recall that the EW-IMP mass $M=1,~2.7,$ and $10~{\rm TeV}$ is
  suggested in order to explain the observed DM abundance when
  $n=2,~3,$ and $5$, respectively, and $Y=0$.  The thermal relic
  abundance of the EW-IMP has not been evaluated when $Y\ne 0$.  We use
  those suggested values in $Y \neq 0$ case since the SI cross section
  is insensitive to the EW-IMP mass as mentioned in the text.  }
\cite{MDM2,Hisano:2006nn}.  It is seen that the cross section is
enhanced as $n$ is larger and $Y$ is smaller. This behavior will be
explained later. We also found that the SI cross section of EW-IMP with
nucleon is far below the current experimental bound. This is the
consequence of the calculation in which all the relevant terms at
leading order are taken into account.  The suppression of the cross
section originates in an accidental cancellation within the SI
effective coupling, $f_N$. Such an cancellation was already pointed
out for the Wino dark matter, {\it i.e.} $n=3$ with $Y=0$
\cite{Hisano:2010fy}. In our calculation, we found that the similar
cancellation also occurs in general $n$-tuplet cases with both
$W$-boson and $Z$-boson contributions.

\begin{figure}
\begin{minipage}{0.5 \hsize}
\begin{center}
\includegraphics[width=80mm]{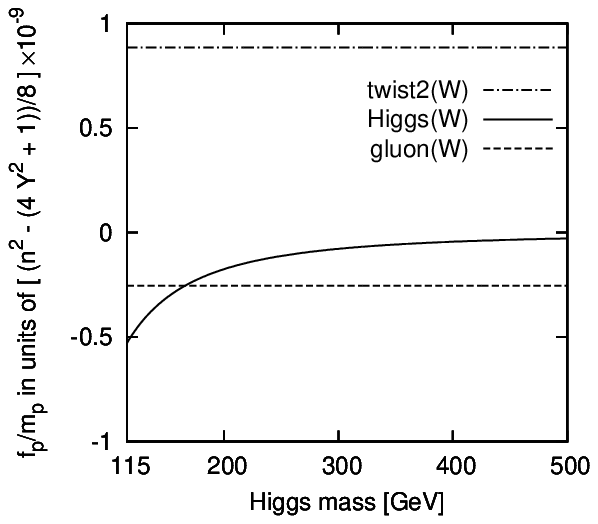}
\end{center}
\end{minipage}
\begin{minipage}{0.5 \hsize}
\begin{center}
\includegraphics[width=80mm]{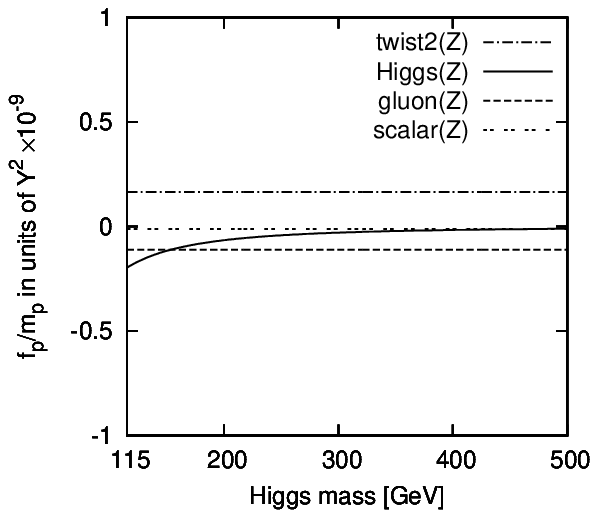}
\end{center}
\end{minipage}
\caption{The contributions of each effective coupling in Eq.~(\ref{f})
  to $f_p /m_p$. Left panel shows each contribution from $W$-boson
  loops as a function of Higgs-boson mass, while right one illustrates
  contribution by $Z$-boson loops. These plots are shown in units of
  $({n^2 - (4 Y^2 + 1)})/{8}$ and $Y^2$, respectively. We take
  $M=10$~TeV.  In the both panels, solid line represents the
  contribution of the Higgs-boson exchange (including both one- and
  two-loop contribution), dashed line indicates the rest of gluon
  contribution in the two-loop contribution, and dash-dotted line
  denotes that from quark twist-2 operators. Double-dashed line in
  right graph shows contribution of scalar-type operator coming from
  box diagrams.}
\label{fig:SIeach}
\end{figure}

Let us examine what happens in the SI effective coupling.  In
Fig.~\ref{fig:SIeach}, we show the contribution to $f_p / m_p$ of each
effective coupling in Eq.~(\ref{f}). The left (right) panel in the
figure shows each contribution due to the $W$-boson ($Z$-boson) loops
as a function of Higgs-boson mass. The plot is given in units of
$({n^2 - (4 Y^2 + 1)})/{8}$ and $Y^2$ in left and right panels,
respectively.  In the figure, we show the contribution of the
Higgs-boson exchange including both one- and two-loop contributions
(solid), the gluon contribution except for the Higgs-boson
contribution (dashed), the quark twist-2 operator contribution
(dash-dotted), and the contribution to the quark scalar-type operators
with the coefficient $f_q$ from the $Z$-boson box diagrams
(double-dashed).  The $W$-boson box diagrams do not generate the
scalar-type operators \cite{Hisano:2010fy}. It is found that the
contributions from the twist-2 operators generated by both $W$-boson
and $Z$-boson loops are positive, while all the other contributions
are negative.  In addition, they have roughly the same order in
absolute value.  This is why the cancellation happens in the SI
effective coupling induced by each $W$-boson and $Z$-boson
contribution.  We also find that the $Z$-boson contribution in units
of $Y^2$ is generally rather small compared with the $W$-boson
contribution in units of $({n^2 - (4 Y^2 + 1)})/{8}$.  Consequently,
the cross section is suppressed when $Y$ gets larger (because the
coupling between $W$ boson and EW-IMP is suppressed).  When Higgs-boson
mass is relatively small as $m_h\sim 115-200~{\rm GeV}$, the
contribution of Higgs-boson exchange becomes large. Then, there is
much more significant cancellation within each $W$- and $Z$-boson
contribution.  As a result, these cancellations give rise to a very
suppressed SI cross section.

\begin{figure}[t]
\begin{minipage}{0.5 \hsize}
\begin{center}
\includegraphics[width=80mm]{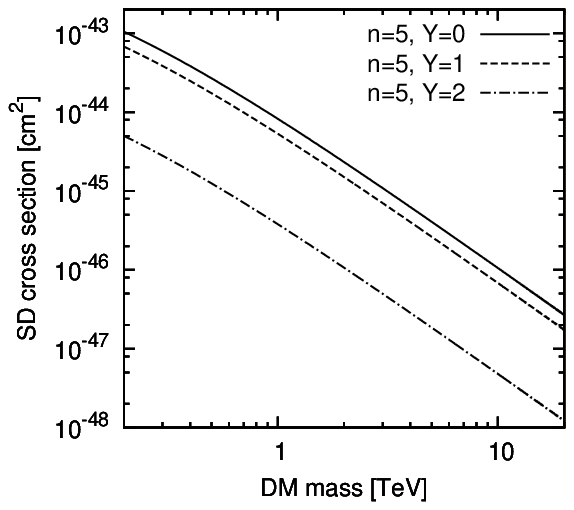}
\end{center}
\end{minipage}
\begin{minipage}{0.5 \hsize}
\begin{center}
\includegraphics[width=80mm]{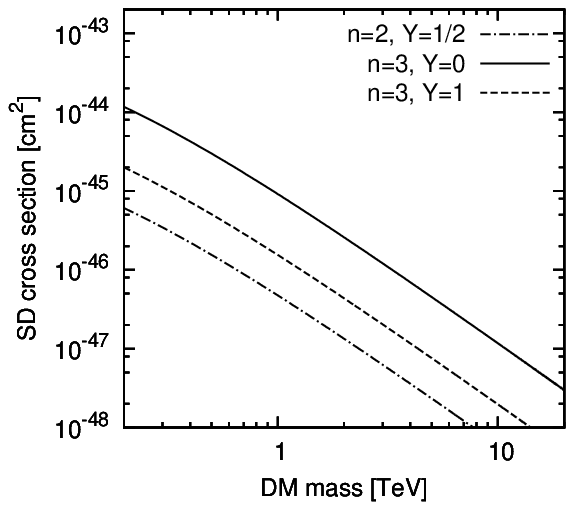}
\end{center}
\end{minipage}
\caption{SD cross section of DM-proton elastic scattering for $n=5$
 (left) and $n=2, 3$ (right). In left graph,  solid line, 
 dashed line, and dash-dotted line represent $Y=0$, 1, 2, while
 in right graph they represents $(n,Y)=(3,0)$, (3,1), (2, 1/2),
 respectively. }
\label{fig:SD}
\end{figure}

Next, we show the SD cross section in Fig.~\ref{fig:SD} for
completeness.  Since the SD cross section is independent of
Higgs-boson mass, we plot it as a function of the EW-IMP mass.
Contrary to the SI cross section, the SD cross section is suppressed
when the EW-IMP mass is larger \cite{Hisano:2004pv}.
In the left panel, the SD cross sections of $n=5$ cases with $Y=0,~1,$
and $2$ are shown in the solid line, the dashed line, and the
dash-dotted line, respectively. In the right panel, we give the cross
sections by taking $(n,Y)=(3,0)$, (3,1) and (2, 1/2), in the solid
line, the dashed line, and the dash-dotted line,
respectively. \footnote{ There was an numerical error in the result of
  SD cross section in Ref.~\cite{Hisano:2010fy}.  We have corrected it
  here.}  As expected, the cross section is enhanced by larger $n$ in
a similar way to the SI cross section. The figure also shows that the
cross section decreases when one sets $Y$ larger, because the
contribution from $W$-boson exchange reduces. We also find that the
contribution from $Z$-boson exchange has the opposite sign and rather
small significance compared to that from $W$-boson loop diagrams.

Throughout the calculation, we have ignored the mass splitting $\delta
m$ between the neutral components for $Y\neq 0$ case. The magnitude of
$\delta m$ is model-dependent.  If $\delta m$ is relatively large
against our assumption, the Higgs-boson exchange contribution at tree
level becomes significant, and one might expect a larger SI cross
section.  Even in such a case, the extension is straightforward.
 
Lastly, we refer to the difference between our results and previous
works.  It is found that the mass functions in
Eq.~(\ref{eq:1loopmassfunc}) in the Appendix \ref{app1} which gives
scalar- and twist-2-type (and also axial-type) contributions agree
with the corresponding loop functions given in Eq.~(42) of
Ref.~\cite{Hisano:2004pv}, except for $F^{(0)}_{S2}(x)$ and
$F^{(0)}_{T1}(x)$ in Ref.~\cite{Hisano:2004pv}. (This was previously
described in Ref.~\cite{Hisano:2010fy}. ) On the other hand, gluon
contribution was taken into account without explicit calculation;
however, the sign of the contribution is the same with our result.  As
a consequence, similarly to our result, the cancellations between each
contribution happen and the cross sections computed in their works are
comparable to our results.

On the other hand, the SI cross sections in Refs.~\cite{MDM,
  Essig:2007az} are larger than our results by more than an order of
magnitude. In Ref.~\cite{MDM} only the scalar-type operator is taken
into account in the limit $M \gg m_W \gg m_q$, which does not agree
with our result in the same limit. Although they mention the twist-2
type contribution, they took into account this contribution in the
same sign as the scalar-type one. Moreover, the gluon contribution is
omitted. Thus, there is no cancellation which we found in the
effective coupling, to give rise to the large cross section.

In Ref.~\cite{Essig:2007az}, the explicit results of the loop
calculation are given. However, all of the loop functions except
$f^Z_I(x)$ in Eq.~(26) in Ref.~\cite{Essig:2007az} are different from
ours. To check their result further, we compare them with ours in the
limit $m_{W/Z}/M \rightarrow 0$. In Eq.~(\ref{eq:higgs}), which is
scalar-type contribution, the first term (of the first parenthesis)
agree with their result except for the sign, and the other terms are
$-1/2$ of corresponding term in their work. For twist-2 type
contributions, the one ({\it i.e.}, Eq.~(\ref{eq:twist2-1})) is the
same as their result in the limit, while another contribution ({\it
  i.e.}, Eq.~(\ref{eq:twist2-2})) is neglected in their paper. The
gluon contribution is considered, but not evaluated properly. As a
result, the cancellation we found in our calculation does not occur,
which makes the resultant cross section large.

\section{Conclusion}
\label{sec:conclusion}

We have studied the direct detection of EW-IMP DM, which is the
lightest neutral component of an $SU(2)_L$ multiplet interacting with
quarks and leptons only through the $SU(2)_L\times U(1)_Y$ gauge
interactions. Although EW-IMP DM does not scatter off nucleon at tree
level, it does at loop level, and the SI scattering cross section is
not suppressed even if the mass of EW-IMP is much larger than gauge
boson mass. We evaluate the two-loop processes for the EW-IMP-gluon
interaction in addition to the one-loop processes of EW-IMP scattering
with light quarks, since both of them yield considerable contribution
to the SI scattering cross section. As a result, the SI cross section
is found to be ${\mathcal O}(10^{-(46-48)}) ~{\rm cm}^2$, depending on
Higgs-boson mass, the number of components $n$ in the multiplet, and
the hypercharge $Y$ of the EW-IMP DM. Such small value of the cross
section is due to an accidental cancellation in the SI effective
coupling.  This cancellation is a general feature for EW-IMP DM, and
thus makes it difficult to catch EW-IMP DM in the direct detection
experiments in the near future.

\section*{Acknowledgments}
This work is supported by Grant-in-Aid for Scientific research from
the Ministry of Education, Science, Sports, and Culture (MEXT), Japan,
No. 20244037, No. 20540252, and No. 22244021 (J.H.), and also by World
Premier International Research Center Initiative (WPI Initiative),
MEXT, Japan. This work was also supported in part by the
U.S. Department of Energy under contract No. DE-FG02-92ER40701, and by
the Gordon and Betty Moore Foundation (K.I.).

\section*{Appendix}
\appendix

In this Appendix we provide the mass functions and the $Z$-boson vacuum
polarization functions, both presented in Section~\ref{sec:results}. 

\section{Mass Functions from One-loop Diagrams}
\label{app1}

The mass functions used in Eqs.~(\ref{eq:higgs}-\ref{eq:twist2-2}) in the
calculation of one-loop diagrams are
\begin{eqnarray}
g_{\rm H}(x)
&=&
-\frac{2}{b_x}
(2+2x-x^2)\tan^{-1} \left(\frac{2 b_x}{\sqrt{x}} \right)
+
2\sqrt{x}(2-x \log(x)) \ ,
\nonumber\\
g_{\text{S}}(x)
&=& \frac{1}{4 b_x} (4-2x+x^2)\tan^{-1} 
\left(\frac{2b_x}{\sqrt{x}}\right) +
\frac{1}{4} \sqrt{x} (2 - x \log (x)) 
\ , \nonumber \\
g_{\rm AV}(x)
&=&
\frac{1}{24b_x}\sqrt{x}(8-x-x^2)\tan^{-1} \left( \frac{2b_x}{\sqrt{x}} \right)
-\frac{1}{24} x(2-(3+x)\log(x)) \ ,
\label{gav}
\nonumber\\
g_{\rm T1}(x)
&=&
\frac{1}{3}b_x(2+x^2)\tan^{-1} \left( \frac{2 b_x}{\sqrt{x}} \right)
+\frac{1}{12}\sqrt{x}(1-2x-x(2-x)\log({x})) \ , 
\nonumber\\
g_{\rm T2}(x)
&=&
\frac{1}{4b_x} x (2-4x+x^2)\tan^{-1} \left( \frac{2 b_x}{\sqrt{x}} \right)
-\frac{1}{4}\sqrt{x}(1-2x-x(2-x)\log(x)) \ ,
\label{eq:1loopmassfunc}
\end{eqnarray}
with ${b_x}\equiv\sqrt{1-x/4}$. 

\section{Mass functions from Two-loop Diagrams}
\label{app2}

Here, we present the polarization functions and the mass functions for
$Z$ boson.
Performing the integration in Eq.(\ref{pol_bc}), we derive the
polarization functions as
\begin{equation}
 \begin{split}
   \left[ \Pi^Z_T (p^2) \right]_{(b)}^q|_{GG} &= \frac{\alpha_2
  \alpha_s}{6 \cos^2 \theta_W}(G^a_{\mu \nu})^2 \left[ (a^V_q)^2
+(a^A_q)^2   \right] \\
&\times\frac{2(p^2-4m_q^2)(p^2-3m_q^2)+4m_q^2(p^2-6m_q^2)
\sqrt{\frac{4m_q^2}{p^2}-1}\cot^{-1}
\left(\sqrt{\frac{4m_q^2}{p^2}-1}\right)}{p^2(p^2-4m_q^2)^2}~, 
\end{split}
\nonumber
\end{equation}
{\footnotesize
\begin{equation}
\begin{split}
&\left[ \Pi^Z_T (p^2) \right]_{(c)}^q|_{GG} 
= -\frac{\alpha_2
  \alpha_s}{3\cos^2 \theta_W}(G^a_{\mu \nu})^2
\Biggl[\frac{  (a^V_q)^2 \left(2 p^4-11
   m_q^2 p^2 +24 m_q^4\right)-(a^A_q)^2  (p^2 -4
   m_q^2) (2 p^2 -3 m_q^2)}{p^2
   (p^2 -4 m_q^2)^2}+ \\
&\frac{2 m_q^2 \sqrt{\frac{4m_q^2}{p^2}-1} \left\{\left(p^4-22
   m_q^2 p^2 +48 m_q^4\right) (a^V_q)^2+(a^A_q)^2  (p^2-4
   m_q^2) (p^2 +6 m_q^2)\right\}\cot
   ^{-1}\left(\sqrt{\frac{4 m_q^2}{p^2}-1}\right)}{p^2
   (p^2 -4 m_q^2)^3}\Biggr] ~,
 \end{split}
\end{equation} 
}
and in the limit of zero quark mass, these functions lead to
\begin{equation}
 \begin{split}
  \left[ \Pi^Z_T (p^2) \right]_{(b)}^q|_{GG}&\rightarrow
\frac{\alpha_2
  \alpha_s}{3 \cos^2 \theta_W}(G^a_{\mu \nu})^2 \left[ (a^V_q)^2
+(a^A_q)^2   \right]\frac{1}{p^2}~, \\
\left[ \Pi^Z_T (p^2) \right]_{(c)}^q|_{GG}&\rightarrow
-\frac{2\alpha_2
  \alpha_s}{3 \cos^2 \theta_W}(G^a_{\mu \nu})^2 \left[ (a^V_q)^2
-(a^A_q)^2   \right]\frac{1}{p^2}~.
 \end{split}
\end{equation}
Using the polarization functions, we compute $g_{\rm Z}(z,y)$ in
Eq.(\ref{fgbc}) as
\begin{equation}
 g_{\rm Z}(z, y)=\left[\sum_{q=u,d,s,c,b}\left\{
(a^V_q)^2+(a^A_q)^2
\right\}-2\sum_{Q=c,b}c_Q\left\{
(a^V_q)^2-(a^A_q)^2
\right\}\right]\times 4g_{\rm B1}(z)+g_t(z,y)~.
\end{equation}
Here, the first term comes from the contribution of all quarks except 
top quark, and $c_Q$ in it represents the QCD corrections in the long-distance
contributions \cite{Djouadi2000}. The second term $g_t(z,y)$ is
the contribution of top quark. The function $g_{\rm B1}(x)$ is given as
\begin{equation}
  g_{\rm B1}(x)=
-\frac{1}{24} \sqrt{x} (x \log
   (x)-2)
+\frac{(x^2-2x
   +4) \tan ^{-1}(\frac{2 b_x}{\sqrt{x}})}{24b_x}~,
\label{gb1}
\end{equation}
which is equal to the one in Ref.~\cite{Hisano:2010fy}.
In the calculation of $g_t(z,y)$, we decompose it into two parts:
\begin{equation}
\begin{split}
g_t(z,y) =  g_{t}^{\text{no-log}} (z,y)+ g_t^{\text{log}}(z,y) .
\end{split}
\end{equation}
$g_{t}^{\text{no-log}} (z,y)$ is analytically obtained as
\begin{equation}
\begin{split}
g_{t}^{\text{no-log}} (z,y)
&= (a^V_t)^2 ~G_{t1}(z,y) + (a^A_t)^2~ G_{t2}(z,y)  ,\\
\end{split}
\end{equation}
where
\begin{equation}
\begin{split}
G_{t1} (z, y )
&= - \frac{\sqrt{z} (12 y^2 - z y + z^2)}{3 (4 y -z)^2} \\
&+ \frac{z^{3/2} ( 48 y^3 -  20 z y^2 + 12 z^2 y - z^3)}{6 (4 y
 -z)^3}~\log z + \frac{2 z^{3/2} y^2 (4 y - 7 z)}{3 (4 y
 -z)^3}~\log (4 y) \\ 
& - \frac{z^{3/2} \sqrt{y} (16 y^3 - 4 (2+ 7z) y^2 + 14 (2 + z) y + 5
 z)}{3 (4 y -z)^3 \sqrt{1 - y}}~\mathrm{tan}^{-1} \left(
 \frac{\sqrt{1-y}}{\sqrt{y}} \right) \\ 
&- \mathrm{tan}^{-1} \left( \frac{\sqrt{4-z}}{\sqrt{z}} \right) \\
&~\times  \frac{48 (z^2 -2 z + 4) y^3 - 4 z (5 z^2 -10 z + 44) y^2 + 12
 z^3 (z-2) y - z^3 (z^2 - 2 z + 4)}{3 (4 y -z)^3 \sqrt{4 - z}} ,  
\nonumber
\end{split}
\end{equation}
\begin{equation}
\begin{split}
G_{t2} (z, y )
&= \frac{\sqrt{z} (2 y - z)}{ (4y -z)} - \frac{z^{3/2} (8 y^2 - 8 z y +
 z^2)}{2 (4 y - z)^2} ~\log z - \frac{4 z^{3/2} y^2}{(4 y -z)^2}
 ~\log (4 y) \\ 
&~~~~~ + \frac{4 z^{3/2} \sqrt{y} (2 y^2 -y -1)}{(4 y -z)^2 \sqrt{1 -
 y}}~\mathrm{tan}^{-1} \left( \frac{\sqrt{1-y}}{\sqrt{y}} \right) \\ 
&~~~~~~~ - \frac{8 z (z^2- 2 z +1) y - (z^2 - 2 z + 4) (8 y^2 + z^2)}{(4
 y -z)^2 \sqrt{4 -z}} ~\mathrm{tan}^{-1} \left(
 \frac{\sqrt{4-z}}{\sqrt{z}} \right) . \\ 
\end{split}
\end{equation}
On the other hand,  we calculate  $g_t^{\text{log}}(z,y)$ numerically.
For convenience, we rewrite it as
\begin{equation}
\begin{split}
g_{t}^{\text{log}} (z,y)
&= 4 z^{3/2} y^2 \left( A_1~y ~[I_1 + I_2 ] + A_2~[I_3 + I_4] \right),
\end{split}
\end{equation}
with
\begin{equation}
\begin{split}
&A_1 = - 2 (a^V_t)^2 + 4 (a^A_t)^2 ,\\
&A_2 = - (a^V_t)^2 + (a^A_t)^2 , \\
\end{split}
\end{equation}
and then carry out the following integrals numerically,
\begin{equation}
\begin{split}
I_1 &=  \int_{0}^{\infty} \mathrm{d}t ~ \frac{(\sqrt{t+4} - \sqrt{t})
 \left( \log\left[ \sqrt{t + 4 y } + \sqrt{t} \right]  -
 \log\left[ \sqrt{t + 4 y} - \sqrt{t} \right]  \right) } {[t+
 z]^2~[t + 4 y]^{5/2}~t}  ,\\ 
\nonumber
%
%
I_2 &=  \int_{0}^{\infty} \mathrm{d}t ~ \frac{1}{2} \times \frac{( t + 2
 - \sqrt{t} \sqrt{t + 4} ) \left( \log\left[ \sqrt{t + 4 y } +
 \sqrt{t} \right]  -  \log\left[ \sqrt{t + 4 y} - \sqrt{t}
 \right]  \right) } {[t+ z]^2~[t + 4 y]^{5/2}~t^{1/2}}  ,\\ 
\nonumber
%
%
I_3 &=  \int_{0}^{\infty} \mathrm{d}t ~ \frac{(\sqrt{t+4} - \sqrt{t})
 \left( \log\left[ \sqrt{t + 4 y } + \sqrt{t} \right]  -
 \log\left[ \sqrt{t + 4 y} - \sqrt{t} \right]  \right) } {[t+
 z]^2~[t + 4 y]^{5/2}} , \\ 
\nonumber
%
%
I_4 &=  \int_{0}^{\infty} \mathrm{d}t ~ \frac{1}{2} \times \frac{
 \sqrt{t} ( t + 2 - \sqrt{t} \sqrt{t + 4} ) \left( \log\left[
 \sqrt{t + 4 y } + \sqrt{t} \right]  -  \log\left[ \sqrt{t + 4 y}
 - \sqrt{t} \right]  \right) } {[t+ z]^2~[t + 4 y]^{5/2}} . \\ 
\end{split}
\end{equation}

Lastly, we present the mass function $g_{\rm W}(w, y)$ in
Eq.~(\ref{fgbc}), which is readily obtained by following the similar
procedure described in Ref.~\cite{Hisano:2010fy}: 
\begin{eqnarray}
g_{\rm W}(w,y)=2 g_{\rm B1}(w) + g_{\rm B3}(w,y)
\label{eq:gw}
\end{eqnarray}
with 
\begin{equation}
g_{\rm B3}(x, y)  
= g^{(1)}_{\rm B3}(x, y)   + c_b g^{(2)}_{\rm B3}(x, y)  .
\label{gb3}
\end{equation}
Here the first term of $g_{\rm W}(w,y)$ is coming from the first- and 
second-generation quark loop diagrams, while the second term is from the
third-generation quark loop diagrams. 
The mass function $g_{\rm B1}(x)$ is displayed in
Eq.~(\ref{gb1}). Although we use the same symbol for $g_{\rm B3}(x,
y)$ as in Ref.~\cite{Hisano:2010fy}, it is reevaluated with the QCD
correction for the long-distance 
contributions, which we illustrate with the factor $c_b$ in Eq.~(\ref{gb3})
explicitly. We have checked numerically that including the QCD
correction changes the SI cross section by up to a  few \%. 
The functions 
$g^{(1)}_{\rm B3}(x, y)$ and $g^{(2)}_{\rm B3}(x, y)$ are analytically given
as
\begin{equation}
\begin{split}
g^{(1)}_{\rm B 3}(x, y)  
&=  \frac{- x^{3/2} }{12 (y - x)} + \frac{- x^{3/2} y^2}{24 (y - x )^2}
 \log y - \frac{x^{5/2} (x- 2 y)}{24 (y - x)^2} \log x
 \\ 
&~ - \frac{x^{3/2} \sqrt{y} (y + 2) \sqrt{4 - y}}{12 (y - x)^2}
 ~\mathrm{tan}^{-1} \left( \frac{\sqrt{4- y} }{\sqrt{y}} \right) \\ 
&~+ \frac{ x  (x^3 -2 (y+ 1) x^2 + 4 (y+1)  x  + 4 y ) }{12 (y - x)^2~
 \sqrt{4 - x}}~\mathrm{tan}^{-1} \left( \frac{\sqrt{4 - x}}{\sqrt{x}}
 \right) , \\
g^{(2)}_{\rm B 3}(x, y)  
&=  \frac{- x^{3/2} y}{12 (y - x)^2} + \frac{- x^{5/2} y^2}{24 (y - x
 )^3} \log y + \frac{x^{5/2} y^2}{24 (y - x)^3} \log x
 \\ 
&~ + \frac{x^{3/2} \sqrt{y} (- 6 y + x y^2 - 2 x y - 2 x )}{12 (y - x)^3
 ~\sqrt{4 - y}} ~\mathrm{tan}^{-1} \left( \frac{\sqrt{4-y}}{\sqrt{y}}
 \right) \\ 
&~+ \frac{- x y  (x^2 y  - 2 x y - 6x - 2y) }{12 (y - x)^3~
 \sqrt{4-x}}~\mathrm{tan}^{-1} \left( \frac{\sqrt{4-x}}{\sqrt{x}}
 \right) ,
\end{split}
\end{equation}
respectively.

{}

\end{document}